\documentclass[showpacs,aps,graphicx]{revtex4}
\usepackage{graphicx}

\begin{document}

\title{Optimal multi-photon entanglement concentration with the photonic Faraday rotation}

\author{Lan Zhou$^{1,2}$ }
\address{$^1$ College of Mathematics \& Physics, Nanjing University of Posts and Telecommunications, Nanjing,
210003, China\\
$^2$ Institute of Signal Processing  Transmission, Nanjing
University of Posts and Telecommunications, Nanjing, 210003,  China\\}

\begin{abstract}
A recent paper (Phys. Rev. A \textbf{86}, 034305 (2012)) proposed an entanglement concentration protocol (ECP) for distilling one pair of maximally entangled multi-photon Greenberger-Horne-Zeilinger (GHZ) state from two pairs of less-entangled multi-photon states with the photonic Faraday rotation. In this paper, we put forward an improved ECP for arbitrary less-entangled multi-photon GHZ state. In the ECP, we only require one pair of less-entangled multi-photon state and one auxiliary photon, and the whole concentration process only requires local operations. Moreover, our ECP can be used repeatedly to further concentrate the less-entangled multi-photon state and obtain a higher success probability. If consider the practical operation and imperfect detection, our protocol is more efficient. This ECP may be useful in current quantum information processing.
\end{abstract}
\date{\today}
\maketitle

\section{introduction}

In recent few decades, the quantum information technology has developed with extremely high speed. Experimental results demonstrate that the quantum communication and computation are much more efficient than their classical counterparts in many aspects. In almost all the practical quantum communication and computation tasks, the maximally entangled quantum states may be the most important resources \cite{rmp,teleportation,cteleportation,densecoding,densecoding1,densecoding2,QSDC,QSDC1,QSDC2,Ekert91,QKDdeng1,QKDdeng2,QSTS,QSS1,QSS2,QSS3}. For example, in the fields of quantum teleportation \cite{teleportation,cteleportation}, quantum dense coding \cite{densecoding1,densecoding2}, quantum communication \cite{QSDC,QSDC1,QSDC2}, and entanglement-based quantum key distribution \cite{Ekert91,QKDdeng1,QKDdeng2}, one needs to use the maximally entangled states to set up the quantum entanglement channel. However, the quality of the maximally entangled states are easily degraded, for the entangled particles may inevitably interact with the channel noise during the storage and transmission process. The maximally entangled state may degrade to the mixed state or less-entangled pure state, which can not set up the high quality quantum entanglement channel \cite{memory}. Therefore, in practical applications, we need to recover the mixed states or the less-entangled pure states into the maximally entangled ones.

Entanglement concentration is an effective method for extracting the maximally entangled quantum states from less-entangled pure states \cite{C.H.Bennett1,Pan1}. In 1996, Bennett \emph{et al.} proposed the first entanglement concentration protocol (ECP), which is called the Schmidt projection method \cite{C.H.Bennett2}. The ECP requires some collective and nondestructive measurements, which are difficult to realize under current experimental condition. After that, various ECPs have been put forward, successively \cite{swapping1,swapping2,Yamamoto1,zhao1,shengpra2,shengqic,shengpla,dengpra,shengpra3,shengpra4,wangxb,wangc,Pengpra}. For example, in 1999, Bose \emph{et al.} proposed an ECP with entanglement swapping \cite{swapping1}, which was improved by Shi \emph{et al.} in 2000 \cite{swapping2}. In 2001, Zhao \emph{et al.} and Yamamoto \emph{et al.} proposed two similar ECPs, independently \cite{zhao1,Yamamoto1}, in which the optical polarization beam splitters (PBSs) are adopted to complete the parity check measurement for photons. In 2008, Sheng \emph{et al.} adopt the cross-Kerr nonlinearity to construct the quantum nondemolition detector, which makes the ECP can be used repeatedly to further concentrate the less-entangled state. In 2010, the ECP for single-photon entanglement was also proposed by Sheng \emph{et al.} \cite{shengqic}. So far, most ECPs have focused on the single-photon and two-photon systems. Only a few ECPs deal with the multi-photon system, for its high operational complexity,.

Recently, the cavity quantum electrodynamics (QED) have been proved to be a powerful platform for performing the QIP tasks of the photon-atom states, due to the controllable interaction between atoms and photons \cite{cavity,cavity1,cavity2,cavity3,cavity4,cavity5,cavity6,cavity7,cavity8,cavity9}. With the atoms strongly interacting with local cavities, the spatially separated cavities could serve as quantum nodes, and construct a quantum network assisted by the photons acting as a quantum bus \cite{cavity6,node,node1,node2}. Especially, during the past few years, available techniques have achieved the input-output process relevant to optical low-quality (Q) cavities, such as the microtoroidal resonator (MTR) \cite{MTR}. It is attractive and applicable to combine the input-output process with low-Q cavities, for it can accomplish high-quality QIP tasks with currently available techniques. In 2009, An \emph{et al.} showed that when a photon interacts with an atom trapped in a low-Q cavity, different polarization of the input photon can cause different phase rotation of the output photon, which is called the photonic Faraday rotation \cite{r}. The photonic Faraday rotation only works in low-Q cavities and is insensitive to both cavity decay and atomic spontaneous emission.
In 2012, based on the photonic Faraday rotation, Peng \emph{et al.} proposed an ECP for the multi-photon Greenberger-Horne-Zeilinger (GHZ) state \cite{Pengpra}, in which they can distill one pair of maximally entangled multi-photon GHZ state from two pairs of less-entangled pure multi-photon GHZ states with some probability. However, this ECP is not optimal, for after the concentration, one pair of less-entangled multi-photon state has to be wasted. In this paper, we will put forward an improved ECP for less-entangled multi-photon GHZ state. In our protocol, we only require one pair of less-entangled multi-photon state, an auxiliary photon and a three-level atom trapped in the low-Q cavity. With the help of the photonic Faraday rotation, we can successfully distill the maximally entangled multi-photon GHZ state with the same success probability as Ref. \cite{Pengpra}. Moreover, our ECP can be used repeatedly to further concentrate the discarded items in Ref. \cite{Pengpra} and obtain a higher success probability. Especially, if we consider the practical operation and imperfect detection, our ECP is more powerful.

This paper is organized as follows: In Sec. II, we first explain the basic principle of the photonic Faraday rotation. In Sec. III, we explain our ECP for the less-entangled two-photon state. In Sec. IV, we extend this ECP to concentrate the N-photon GHZ state. In Sec. V, we make a discussion and summary.

\section{Photonic Faraday rotation}
\begin{figure}[!h]
\begin{center}
\includegraphics[width=8cm,angle=0]{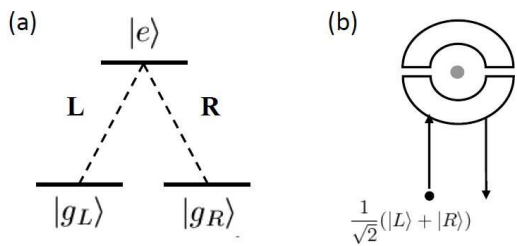}
\caption{A schematic drawing of the interaction between the photon pulse and the three-level atom in the low-Q cavity. (a) the three-level atom trapped in the low-Q cavity. $|g_{L}\rangle$ and $|g_{R}\rangle$ represent two Zeeman sublevels of its degenerate ground state, and $|e\rangle$ represents its excited state. (b) The interaction between the photon pulse and the three-level atom in the low-Q cavity. The state $|g_{L}\rangle$ and $|g_{R}\rangle$ couple with a left (L) polarized and a right (R) polarized photon, respectively.}
\end{center}
\end{figure}
The photonic Faraday rotation is the key element of our ECP. In this section, we will briefly explain the basic input-output relation for a single photon pulse coherently interacting with a trapped three-level atom. As shown in Fig. 1, a three-level atom is trapped in a low-Q cavity, where the state $|e\rangle$ represents the excited state, and the states $|g_{L}\rangle$ and $|g_{R}\rangle$ represent two Zeeman sublevels of the ground state, respectively. A single photon pulse with the form $|\varphi_{in}\rangle=\frac{1}{\sqrt{2}}(|L\rangle+|R\rangle)$ enters the low-Q cavity and interacts coherently with the three-level atom. The $|L\rangle$ and $|R\rangle$ represent the left-circularly polarization and right-circularly polarization of the input photon, respectively. During the interaction process, the atom will achieve the transition $|g_{L}\rangle\leftrightarrow|e\rangle$ ($|g_{R}\rangle\leftrightarrow|e\rangle$) by absorbing or emitting a $|L\rangle$ ($|R\rangle$) circularly polarized photon.

Based on the research from Refs. \cite{Faraday,faraday3,faraday1}, if we consider the low-Q cavity limit and the
weak excitation limit, we can solve the Langevin equations of the motion for the cavity and atom analytically. In this way, we can draw a general expression for the single relation between the input ($a_{in,j(t)}$) and output ($a_{out,j(t)}$) single-photon state in the form\cite{Pengpra,Faraday,faraday3,faraday1,r}
\begin{eqnarray}
r(\omega_{p})\equiv\frac{a_{out,j(t)}}{a_{in,j(t)}}=\frac{[i(\omega_{c}-\omega_{p})-\frac{\kappa}{2}][i(\omega_{0}-\omega_{p})+\frac{\gamma}{2}]+g^{2}}
{[i(\omega_{c}-\omega_{p})+\frac{\kappa}{2}][i(\omega_{0}-\omega_{p})+\frac{\gamma}{2}]+g^{2}}.\label{r}
\end{eqnarray}
In Eq. (\ref{r}), $\kappa$ and $\gamma$ are the cavity damping rate and atomic decay rate, respectively. $\omega_{c}$ is the atomic frequency, $\omega_{p}$ is the input photonic frequency, and $g$ is the atom-cavity coupling strength. Particularly, in the case that the atom uncouples to the cavity, which makes $g=0$, Eq. (\ref{r}) can be simplied as
\begin{eqnarray}
r_{0}(\omega_{p})=\frac{i(\omega_{c}-\omega_{p})-\frac{\kappa}{2}}{i(\omega_{c}-\omega_{p})+\frac{\kappa}{2}}.\label{r0}
\end{eqnarray}
It is obvious that Eq. (\ref{r0}) can be rewritten as a pure phase shift as $r_{0}(\omega_{p})=e^{i\phi_{0}}$. It indicates that if the single-photon pulse senses the empty cavity, the output photon state will convert to $|\varphi_{out}\rangle=r_{0}(\omega_{p})|L (R)\rangle= e^{i\phi_{0}}|L (R)\rangle$. On the other hand, when the single photon couples to the three-level atom, as the photon experiences an extremely weak absorption during the photon-atom interaction process, we can consider that the output photon only experiences a pure phase shift without any absorption for a good approximation. In this way, with strong $\kappa$, weak $\gamma$ and $g$, Eq. (\ref{r}) can also be rewritten as a pure phase shift $r(\omega_{p})\approx e^{i\phi}$ and the output photon state will convert to $|\varphi_{out}\rangle=r(\omega_{p})|L (R)\rangle\approx e^{i\phi}|L (R)\rangle$. Therefore, for an input single-photon state as $|\varphi_{in}\rangle=\frac{1}{\sqrt{2}}(|L\rangle+|R\rangle)$, if the initial atom state is $|g_{L}\rangle$, the output photon state will convert to
\begin{eqnarray}
|\varphi_{out}\rangle_{-}=\frac{1}{\sqrt{2}}(e^{i\phi}|L\rangle+e^{i\phi_{0}}|R\rangle),\label{L}
\end{eqnarray}
while if the initial atom state is $|g_{R}\rangle$, the output photon state will convert to
\begin{eqnarray}
|\varphi_{out}\rangle_{+}=\frac{1}{\sqrt{2}}(e^{i\phi_{0}}|L\rangle+e^{i\phi}|R\rangle).\label{R}
\end{eqnarray}

It can be found that after passing through the low-Q cavity, the polarization direction of the output photon rotates an angle  $\Theta^{-}_{F}=\frac{\phi_{0}-\phi}{2}$ or $\Theta^{+}_{F}=\frac{\phi-\phi_{0}}{2}$, which is so called the photonic Faraday rotation.

According to Eq. (\ref{r}) and Eq. (\ref{r0}), it can be seen that under suitable case, where $\omega_{0}=\omega_{c}$, $\omega_{p}=\omega_{c}-\frac{\kappa}{2}$, and $g=\frac{\kappa}{2}$, we can get $\phi=\pi$ and $\phi_{0}=\frac{\pi}{2}$, so that the relation between the input and output photon can be simplified as \cite{Pengpra}
\begin{eqnarray}
&&|L\rangle|g_{L}\rangle\rightarrow -|L\rangle|g_{L}\rangle,\qquad |R\rangle|g_{L}\rangle\rightarrow i|R\rangle|g_{L}\rangle,\nonumber\\
&&|L\rangle|g_{R}\rangle\rightarrow i|L\rangle|g_{R}\rangle,\qquad |R\rangle|g_{R}\rangle\rightarrow -|R\rangle|g_{R}\rangle.\label{rule}
\end{eqnarray}
Based on the input-output relation of the single photon as Eq. (\ref{rule}), we can perform the entanglement concentration for the arbitrary less-entangled multi-photon GHZ state.

\section{The ECP for the less-entangled two-photon state}
\begin{figure}[!h]
\begin{center}
\includegraphics[width=8cm,angle=0]{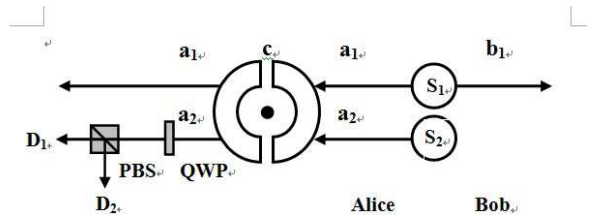}
\caption{A schematic drawing of our ECP for the pure less-entangled two-photon state. S$_{1}$ is the less entanglement source, and S$_{2}$ is the single photon source. We suppose Alice and Bob share a less-entangled two-photon state in the spatial mode a$_{1}$ and b$_{1}$. S$_{2}$ emits an auxiliary photon with the form $|\phi\rangle=\alpha|L\rangle+\beta|R\rangle$, and sends it to Alice in the spatial mode a$_{2}$. Alice makes the photons in the modes a$_{1}$ and a$_{2}$ pass through the low-Q cavity c, successively. After the interaction between the photons and the atom, by measuring the states of the three-level atom and the output auxiliary photon, Alice can successfully distill the maximally entangled two-photon state. Our ECP can be used repeatedly to further concentrate the less-entangled two-photon state.}
\end{center}
\end{figure}
Now, we begin to explain our ECP for the arbitrary less-entangled two-photon state. We suppose that Alice and Bob share a less-entangled pure two-photon state from the less entanglement source S$_{1}$ with the form as
\begin{eqnarray}
|\phi\rangle_{a_{1}b_{1}}=\alpha|L\rangle_{a_{1}}|R\rangle_{b_{1}}+\beta|R\rangle_{a_{1}}|L\rangle_{b_{1}},\label{initial}
\end{eqnarray}
where $|L\rangle$ and $|R\rangle$ represent the left-circularly polarization and right-circularly polarization of the photons, respectively. $\alpha$, $\beta$ are the initial entanglement coefficients, where $|\alpha|^{2}+|\beta|^{2}=1$. As shown in Fig. 2, the single photon source S$_{2}$ emits an auxiliary single photon and sends it to Alice in the spatial mode a$_{2}$. The auxiliary single photon state can be written as
 \begin{eqnarray}
|\phi\rangle_{a_{2}}=\alpha|L\rangle_{a_{2}}+\beta|R\rangle_{a_{2}}.\label{photon}
\end{eqnarray}

Then, Alice introduces a three-level atom trapped in a low-Q cavity c. The initial state of the three-level atom can be described as
\begin{eqnarray}
|\varphi\rangle=\frac{1}{\sqrt{2}}(|g_{L}\rangle+|g_{R}\rangle).
\end{eqnarray}
In this way, the whole three-photon system combined with the three-level atom can be described as
\begin{eqnarray}
|\Phi\rangle&=&|\phi\rangle_{a_{1}b_{1}}\otimes|\phi\rangle_{a_{2}}\otimes|\varphi\rangle=\alpha^{2}|L\rangle_{a_{1}}|R\rangle_{b_{1}}|L\rangle_{a_{2}}|g_{L}\rangle+
\alpha^{2}|L\rangle_{a_{1}}|R\rangle_{b_{1}}|L\rangle_{a_{2}}|g_{R}\rangle\nonumber\\
&+&\alpha\beta|L\rangle_{a_{1}}|R\rangle_{b_{1}}|R\rangle_{a_{2}}|g_{L}\rangle+\alpha\beta|L\rangle_{a_{1}}|R\rangle_{b_{1}}|R\rangle_{a_{2}}|g_{R}\rangle
+\alpha\beta|R\rangle_{a_{1}}|L\rangle_{b_{1}}|L\rangle_{a_{2}}|g_{L}\rangle\nonumber\\
&+&\alpha\beta|R\rangle_{a_{1}}|L\rangle_{b_{1}}|L\rangle_{a_{2}}|g_{R}\rangle+\beta^{2}|R\rangle_{a_{1}}|L\rangle_{b_{1}}|R\rangle_{a_{2}}|g_{L}\rangle
+\beta^{2}|R\rangle_{a_{1}}|L\rangle_{b_{1}}|R\rangle_{a_{2}}|g_{R}\rangle.\label{whole}
\end{eqnarray}

 Alice makes the photons in the spatial modes a$_{1}$ and a$_{2}$ enter the low-Q cavity c, successively. According to the relation between the input and output photon in Eq. (\ref{rule}), Eq. (\ref{whole}) can evolve to
\begin{eqnarray}
|\Phi\rangle_{out1}&=&\alpha^{2}|L\rangle_{a_{1}}|R\rangle_{b_{1}}|L\rangle_{a_{2}}|g_{L}\rangle+(-\alpha^{2})|L\rangle_{a_{1}}|R\rangle_{b_{1}}|L\rangle_{a_{2}}|g_{R}\rangle
+(-i\alpha\beta)|L\rangle_{a_{1}}|R\rangle_{b_{1}}|R\rangle_{a_{2}}|g_{L}\rangle\nonumber\\
&+&(-i\alpha\beta)|L\rangle_{a_{1}}|R\rangle_{b_{1}}|R\rangle_{a_{2}}|g_{R}\rangle
+(-i\alpha\beta)|R\rangle_{a_{1}}|L\rangle_{b_{1}}|L\rangle_{a_{2}}|g_{L}\rangle
+(-i\alpha\beta)|R\rangle_{a_{1}}|L\rangle_{b_{1}}|L\rangle_{a_{2}}|g_{R}\rangle\nonumber\\
&+&(-\beta^{2})|R\rangle_{a_{1}}|L\rangle_{b_{1}}|R\rangle_{a_{2}}|g_{L}\rangle
+\beta^{2}|R\rangle_{a_{1}}|L\rangle_{b_{1}}|R\rangle_{a_{2}}|g_{R}\rangle.\label{out}
\end{eqnarray}

Then, Alice performs the Hadamard (H) operation on the three-level atom and the auxiliary photon in the spatial mode a$_{2}$. The H operation on the atom can be performed by driving it with an external classical field (polarized lasers), which can make
\begin{eqnarray}
|g_{L}\rangle\rightarrow\frac{1}{\sqrt{2}}(|g_{L}\rangle+|g_{R}\rangle), \qquad |g_{R}\rangle\rightarrow\frac{1}{\sqrt{2}}(|g_{L}\rangle-|g_{R}\rangle).\label{hadamard}
\end{eqnarray}
The H operation on the auxiliary photon is performed by making the photon pass through an quarter-wave plate (QWP), which can make
\begin{eqnarray}
|L\rangle_{a_{2}}\rightarrow\frac{1}{\sqrt{2}}(|H\rangle+|V\rangle), \qquad |R\rangle_{a_{2}}\rightarrow\frac{1}{\sqrt{2}}(|H\rangle-|V\rangle),\label{hadamard1}
\end{eqnarray}
where $|H\rangle$ represents the horizontal polarization and $|V\rangle$ represents the vertical polarization.

After the H operations,  Eq. (\ref{out}) can ultimately evolve to
\begin{eqnarray}
|\Phi'\rangle_{out1}&=&\alpha^{2}|L\rangle_{a_{1}}|R\rangle_{b_{1}}|H\rangle|g_{R}\rangle
-\beta^{2}|R\rangle_{a_{1}}|L\rangle_{b_{1}}|H\rangle|g_{R}\rangle
+\alpha^{2}|L\rangle_{a_{1}}|R\rangle_{b_{1}}|V\rangle|g_{R}\rangle\nonumber\\
&+&\beta^{2}|R\rangle_{a_{1}}|L\rangle_{b_{1}}|V\rangle|g_{R}\rangle
+(-i\alpha\beta)|L\rangle_{a_{1}}|R\rangle_{b_{1}}|H\rangle|g_{L}\rangle
+(-i\alpha\beta)|R\rangle_{a_{1}}|L\rangle_{b_{1}}|H\rangle|g_{L}\rangle\nonumber\\
&+&(-i\alpha\beta)|R\rangle_{a_{1}}|L\rangle_{b_{1}}|V\rangle|g_{L}\rangle
-(-i\alpha\beta)|L\rangle_{a_{1}}|R\rangle_{b_{1}}|V\rangle|g_{L}\rangle.\label{whole1}
\end{eqnarray}

Then, Alice makes the output auxiliary photon pass through the PBS, which can transmit the photon in the horizontal polarization $|H\rangle$ and reflect the photon in the vertical polarization $|V\rangle$, respectively. Finally, she measures the state of the three-level atom and the auxiliary photon by the detectors. Based on her measurement results, the Eq. (\ref{whole1})
will collapse to four possible cases. For example, if the measurement results is $|H\rangle|g_{L}\rangle$, Eq. (\ref{whole1}) will collapse to
\begin{eqnarray}
|\phi_{1}\rangle_{a_{1}b_{1}}=\frac{1}{\sqrt{2}}(|L\rangle_{a_{1}}|R\rangle_{b_{1}}+|R\rangle_{a_{1}}|L\rangle_{b_{1}}),\label{max}
\end{eqnarray}
while if the measurement result is $|V\rangle|g_{L}\rangle$, Eq. (\ref{whole1}) will collapse to
\begin{eqnarray}
|\phi'_{1}\rangle_{a_{1}b_{1}}=\frac{1}{\sqrt{2}}(|L\rangle_{a_{1}}|R\rangle_{b_{1}}-|R\rangle_{a_{1}}|L\rangle_{b_{1}}).\label{max1}
\end{eqnarray}
Both Eq. (\ref{max}) and Eq. (\ref{max1}) are the maximally entangled two-photon state, and there is only a phase difference between them. If Eq. (\ref{max1}) is obtained, Alice only needs to perform the phase flip operation with the help of a half-wave plate, Eq. (\ref{max1}) can be converted to Eq. (\ref{max}). So far, our concentration process is completed, where we successfully distill the maximally entangled two-photon state from arbitrary less-entangled photon state, with the success probability of P$=2|\alpha\beta|^{2}$.

On the other hand, there are still two possible measurement results. If the measurement result is $|V\rangle|g_{R}\rangle$,
Eq. (\ref{whole1}) will collapse to
\begin{eqnarray}
|\phi_{2}\rangle_{a_{1}b_{1}}=\alpha^{2}|L\rangle_{a_{1}}|R\rangle_{b_{1}}+\beta^{2}|R\rangle_{a_{1}}|L\rangle_{b_{1}},\label{new}
\end{eqnarray}
while if the measurement result is $|H\rangle|g_{R}\rangle$, Eq. (\ref{whole1}) will collapse to
\begin{eqnarray}
|\phi'_{2}\rangle_{a_{1}b_{1}}=\alpha^{2}|L\rangle_{a_{1}}|R\rangle_{b_{1}}-\beta^{2}|R\rangle_{a_{1}}|L\rangle_{b_{1}}.\label{new1}
\end{eqnarray}
Similar with Eq. (\ref{max}) and Eq. (\ref{max1}), Eq. (\ref{new1}) can be easily converted to Eq. (\ref{new}) by the phase flip operation. Interestingly, it can be found that Eq. (\ref{new}) has the similar form with Eq. (\ref{initial}), that is to say, Eq. (\ref{new}) is a new less-entangled two-photon state. According to the concentration step described above, Eq. (\ref{new}) can be reconcentrated for the next round.

In the second concentration round, S$_{2}$ emits another auxiliary single photon and sends it to Alice in the spatial mode a$_{2}$. The single photon state is with the form
 \begin{eqnarray}
|\phi\rangle_{a_{2}}=\alpha^{2}|L\rangle_{a_{2}}+\beta^{2}|R\rangle_{a_{2}}.\label{photon2}
\end{eqnarray}

Alice makes the photons in the spatial mode a$_{1}$ and a$_{2}$ enter the low-Q cavity, sequentially. Similarly, based on Eq. (\ref{rule}), the whole three-photon system combined with the three-level atom can evolve to
\begin{eqnarray}
|\Phi\rangle_{out2}&=&\alpha^{4}|L\rangle_{a_{1}}|R\rangle_{b_{1}}|L\rangle_{a_{2}}|g_{L}\rangle
+(-\alpha^{4})|L\rangle_{a_{1}}|R\rangle_{b_{1}}|L\rangle_{a_{2}}|g_{R}\rangle
+(-i\alpha^{2}\beta^{2})|L\rangle_{a_{1}}|R\rangle_{b_{1}}|R\rangle_{a_{2}}|g_{L}\rangle\nonumber\\
&+&(-i\alpha^{2}\beta^{2})|L\rangle_{a_{1}}|R\rangle_{b_{1}}|R\rangle_{a_{2}}|g_{R}\rangle
+(-i\alpha^{2}\beta^{2})|R\rangle_{a_{1}}|L\rangle_{b_{1}}|L\rangle_{a_{2}}|g_{L}\rangle
+(-i\alpha^{2}\beta^{2})|R\rangle_{a_{1}}|L\rangle_{b_{1}}|L\rangle_{a_{2}}|g_{R}\rangle\nonumber\\
&+&(-\beta^{4})|R\rangle_{a_{1}}|L\rangle_{b_{1}}|R\rangle_{a_{2}}|g_{L}\rangle
+\beta^{4}|R\rangle_{a_{1}}|L\rangle_{b_{1}}|R\rangle_{a_{2}}|g_{R}\rangle.\label{out2}
\end{eqnarray}

Then, Alice performs the H operation on the atom and the auxiliary single photon, and Eq. (\ref{out2}) can  evolve to
\begin{eqnarray}
|\Phi'\rangle_{out1}&=&\alpha^{4}|L\rangle_{a_{1}}|R\rangle_{b_{1}}|H\rangle|g_{R}\rangle
-\beta^{4}|R\rangle_{a_{1}}|L\rangle_{b_{1}}|H\rangle|g_{R}\rangle
+\alpha^{4}|L\rangle_{a_{1}}|R\rangle_{b_{1}}|V\rangle|g_{R}\rangle\nonumber\\
&+&\beta^{4}|R\rangle_{a_{1}}|L\rangle_{b_{1}}|V\rangle|g_{R}\rangle
+(-i\alpha^{2}\beta^{2})|L\rangle_{a_{1}}|R\rangle_{b_{1}}|H\rangle|g_{L}\rangle
+(-i\alpha^{2}\beta^{2})|R\rangle_{a_{1}}|L\rangle_{b_{1}}|H\rangle|g_{L}\rangle\nonumber\\
&+&(-i\alpha^{2}\beta^{2})|R\rangle_{a_{1}}|L\rangle_{b_{1}}|V\rangle|g_{L}\rangle
-(-i\alpha^{2}\beta^{2})|L\rangle_{a_{1}}|R\rangle_{b_{1}}|V\rangle|g_{L}\rangle.\label{whole2}
\end{eqnarray}

Finally, by detecting the quantum states of the three-level atom and the auxiliary photon, Eq. (\ref{whole2}) can also collapse to four possible cases. If the measurement result is $|H\rangle|g_{L}\rangle$ or $|V\rangle|g_{L}\rangle$, Eq. (\ref{whole2}) will collapse to
\begin{eqnarray}
|\phi_{3}\rangle_{a_{1}b_{1}}=\frac{1}{\sqrt{2}}(|L\rangle_{a_{1}}|R\rangle_{b_{1}}\pm|R\rangle_{a_{1}}|L\rangle_{b_{1}}),\label{max2}
\end{eqnarray}
where $'+'$ corresponds to $|H\rangle|g_{L}\rangle$, and $'-'$ corresponds to $|V\rangle|g_{L}\rangle$. Therefore, we can successfully distill the maximally entangled two-photon state in the second concentration round, with the success probability of P$_{2}=\frac{2|\alpha\beta|^{4}}{|\alpha|^{4}+|\beta|^{4}}$, where the subscript '2' means in the second concentration round. If the measurement result is $|V\rangle|g_{R}\rangle$ or $|H\rangle|g_{R}\rangle$, Eq. (\ref{whole2}) can ultimately collapse to
\begin{eqnarray}
|\phi_{3}\rangle_{a_{1}b_{1}}=\alpha^{4}|L\rangle_{a_{1}}|R\rangle_{b_{1}}\pm\beta^{4}|R\rangle_{a_{1}}|L\rangle_{b_{1}},\label{new2}
\end{eqnarray}
which are new less-entangled two-photon states and can be reconcentrated for the third round.

In this way, we have proved that by providing an auxiliary photon with the form $|\phi\rangle=\alpha^{2^{K-1}}|L\rangle+\beta^{2^{K-1}}|R\rangle$, where 'K' represents the iteration times, our ECP can be used repeatedly to further concentrate the less-entangled photon state. It is worth noting that for preparing the single photon state, we need to know the exact value of the initial entanglement coefficients $\alpha$ and $\beta$ in advance. Actually, some other ECPs also have this requirement\cite{swapping2,shengpra3,shengpra4,dengpra,wangc}. According to the previous research results, the exact value of $\alpha$ and $\beta$ can be obtained by measuring an enough amount of the target samples.

\section{The ECP for the less-entangled N-photon GHZ state}
\begin{figure}[!h]
\begin{center}
\includegraphics[width=8cm,angle=0]{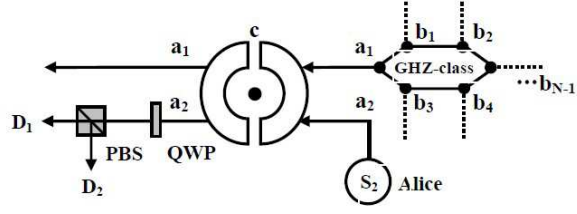}
\caption{A schematic drawing of our ECP for the pure less-entangled N-photon GHZ state. The less-entangled N-photon state are shared by N parties. Alice introduces an auxiliary single photon with the form $|\phi\rangle=\alpha|L\rangle+\beta|R\rangle$ and an three-level atom trapped in the low-Q cavity c. Alice makes photons in the spatial mode a$_{1}$ and a$_{2}$ pass through the low-Q cavity c, successively. By measuring the states of the three-level atom and the output auxiliary photon, we can successfully distill the maximally entangled N-photon GHZ state. Our ECP can also be used repeatedly to further concentrate the less-entangled N-photon GHZ state.}
\end{center}
\end{figure}

Our ECP can be extended to concentrate the less-entangled N-photon GHZ state. We suppose that a less-entangled N-photon state is possessed by N parties, say Alice, Bob, Charlie and so on. The photon in the hand of Alice is in the spatial mode a$_{1}$, while other $N-1$ photons in the hands of other parties are in the spatial modes b$_{1}$, b$_{2}$, $\cdots$, b$_{N-1}$, respectively. The less-entangled N-photon GHZ state can be described as
\begin{eqnarray}
|\phi\rangle_{N}=\alpha|L\rangle_{a_{1}}|R\cdots R\rangle_{b_{1}\cdots b_{N-1}}+\beta|R\rangle_{a_{1}}|L\cdots L\rangle_{b_{1}\cdots b_{N-1}}.\label{initialN}
\end{eqnarray}

Similar with Sec. III, the single photon source S emits an auxiliary photon and sends it to Alice in the spatial mode a$_{2}$ with the form of Eq. (\ref{photon}). Alice makes the photons in the spatial modes a$_{1}$ and a$_{2}$ pass through the low-Q cavity c, successively. After the cavity, the whole $N+1$ photons state combined with the three-level atom state can evolve to
\begin{eqnarray}
|\Phi\rangle_{outN}&=&\alpha^{2}|L\rangle_{a_{1}}|L\rangle_{a_{2}}|R\cdots R\rangle_{b_{1}\cdots b_{N-1}}|g_{L}\rangle
+(-\alpha^{2})|L\rangle_{a_{1}}|L\rangle_{a_{2}}|R\cdots R\rangle_{b_{1}\cdots b_{N-1}}|g_{R}\rangle\nonumber\\
&+&(-i\alpha\beta)|L\rangle_{a_{1}}|R\rangle_{a_{2}}|R\cdots R\rangle_{b_{1}\cdots b_{N-1}}|g_{L}\rangle
+(-i\alpha\beta)|L\rangle_{a_{1}}|R\rangle_{a_{2}}|R\cdots R\rangle_{b_{1}\cdots b_{N-1}}|g_{R}\rangle\nonumber\\
&+&(-i\alpha\beta)|R\rangle_{a_{1}}|L\rangle_{a_{2}}|L\cdots L\rangle_{b_{1}\cdots b_{N-1}}|g_{L}\rangle
+(-i\alpha\beta)|R\rangle_{a_{1}}|L\rangle_{a_{2}}|L\cdots L\rangle_{b_{1}\cdots b_{N-1}}|g_{R}\rangle\nonumber\\
&+&(-\beta^{2})|R\rangle_{a_{1}}|R\rangle_{a_{2}}|L\cdots L\rangle_{b_{1}\cdots b_{N-1}}|g_{L}\rangle
+\beta^{2}|R\rangle_{a_{1}}|R\rangle_{a_{2}}|L\cdots L\rangle_{b_{1}\cdots b_{N-1}}|g_{R}\rangle.\label{outN}
\end{eqnarray}

Here, Alice only needs to perform the H operation on the three-level atom and the auxiliary single photon in the a$_{2}$ mode. After the H operation, Eq. (\ref{outN}) will ultimately evolve to
\begin{eqnarray}
|\Phi'\rangle_{out1}&=&\alpha^{2}|L\rangle_{a_{1}}|R\cdots R\rangle_{b_{1}\cdots b_{N-1}}|H\rangle|g_{R}\rangle
-\beta^{2}|R\rangle_{a_{1}}|L\cdots L\rangle_{b_{1}\cdots b_{N-1}}|H\rangle|g_{R}\rangle\nonumber\\
&+&\alpha^{2}|L\rangle_{a_{1}}|R\cdots R\rangle_{b_{1}\cdots b_{N-1}}|V\rangle|g_{R}\rangle
+\beta^{2}|R\rangle_{a_{1}}|L\cdots L\rangle_{b_{1}\cdots b_{N-1}}|V\rangle|g_{R}\rangle\nonumber\\
&+&(-i\alpha\beta)|L\rangle_{a_{1}}|R\cdots R\rangle_{b_{1}\cdots b_{N-1}}|H\rangle|g_{L}\rangle
+(-i\alpha\beta)|R\rangle_{a_{1}}|L\cdots L\rangle_{b_{1}\cdots b_{N-1}}|H\rangle|g_{L}\rangle\nonumber\\
&+&(-i\alpha\beta)|R\rangle_{a_{1}}|L\cdots L\rangle_{b_{1}\cdots b_{N-1}}|V\rangle|g_{L}\rangle
-(-i\alpha\beta)|L\rangle_{a_{1}}|R\cdots R\rangle_{b_{1}\cdots b_{N-1}}|V\rangle|g_{L}\rangle.\label{wholeN}
\end{eqnarray}

Finally, Alice measures quantum states of the three-level atom and the auxiliary single photon. If the measurement result is $|H\rangle|g_{L}\rangle$,
Eq. (\ref{wholeN}) will collapse to
\begin{eqnarray}
|\phi_{1}\rangle_{N}=\frac{1}{\sqrt{2}}(|L\rangle_{a_{1}}|R\cdots R\rangle_{b_{1}\cdots b_{N-1}}+|R\rangle_{a_{1}}|L\cdots L\rangle_{b_{1}\cdots b_{N-1}}),\label{maxN}
\end{eqnarray}
while if the measurement result is $|V\rangle|g_{L}\rangle$, Eq. (\ref{wholeN}) will collapse to
\begin{eqnarray}
|\phi'_{1}\rangle_{N}=\frac{1}{\sqrt{2}}(|L\rangle_{a_{1}}|R\cdots R\rangle_{b_{1}\cdots b_{N-1}}-|R\rangle_{a_{1}}|L\cdots L\rangle_{b_{1}\cdots b_{N-1}}).\label{max1N}
\end{eqnarray}
Both Eq. (\ref{maxN}) and Eq. (\ref{max1N}) are maximally entangled N-photon GHZ state. Eq. (\ref{max1N}) can be easily converted to Eq. (\ref{maxN}) by the phase flip operation. Therefore, so far, we have successfully recover the less-entangled N-photon GHZ state into the maximally entangled N-photon GHZ state, with the success probability of P$=2|\alpha\beta|^{2}$, which is the same as the success probability in Sec.III.

Meanwhile, there are still another two possible measurement results. If the result is $|H\rangle|g_{R}\rangle$, Eq. (\ref{wholeN}) will collapse to
\begin{eqnarray}
|\phi_{2}\rangle_{N}=\alpha^{2}|L\rangle_{a_{1}}|R\cdots R\rangle_{b_{1}\cdots b_{N-1}}+\beta^{2}|R\rangle_{a_{1}}|L\cdots L\rangle_{b_{1}\cdots b_{N-1}},\label{newN}
\end{eqnarray}
while if the result is $|V\rangle|g_{R}\rangle$, Eq. (\ref{wholeN}) will collapse to
\begin{eqnarray}
|\phi'_{2}\rangle_{N}=\alpha^{2}|L\rangle_{a_{1}}|R\cdots R\rangle_{b_{1}\cdots b_{N-1}}-\beta^{2}|R\rangle_{a_{1}}|L\cdots L\rangle_{b_{1}\cdots b_{N-1}}.\label{new1N}
\end{eqnarray}
Eq. (\ref{new1N}) also can be converted to Eq. (\ref{newN}) by the phase flip operation. It is obvious that Eq. (\ref{newN}) has the similar form with Eq. (\ref{initialN}), that is to say, Eq. (\ref{newN}) is a new less-entangled N-photon GHZ state. Based on the concentration step described above, Alice only needs to introduce an auxiliary single photon with the form of Eq. (\ref{photon2}), Eq. (\ref{newN}) can be reconcentrated for the next round. Therefore, Alice only needs to provide an auxiliary single photon with the form $|\phi\rangle=\alpha^{2^{K-1}}|L\rangle+\beta^{2^{K-1}}|R\rangle$ in each concentration round, our ECP can also be used repeatedly to further concentrate the N-photon GHZ state. In each concentration round, its success probability is the same as that in Sec. III.

\section{discussion and summary}

By far, we have fully explained our ECP for less-entangled multi-photon GHZ state. In our ECP, with the help of the photonic Faraday rotation, we can extract the maximally entangled multi-photon GHZ state from less-entangled multi-photon state with some probability. Comparing with the ECP in Ref. \cite{Pengpra}, our ECP has several obvious advantages. First, the ECP in Ref. \cite{Pengpra} can distill one pair of maximally entangled multi-photon state from two pairs of less-entangled multi-photon states. Our ECP only requires a pair of less-entangled multi-photon state and an auxiliary photon. Therefore, our ECP is more economic. Second, in Ref. \cite{Pengpra}, after successfully performed the ECP, each of the N parties should measure his or her photon state to ultimately obtain the maximally entangled N-photon GHZ state. Finally, they also should check their measurement results to confirm the remained maximally entangled state. Therefore, it increases the operational complexity greatly, especially when the photon number N is large. Our ECP only requires local operations. During the concentration process, Alice only needs to detect the atom state and the auxiliary photon state in each concentration round, and finally tells the result to the other parties, which can simplify the operation largely. Third, our ECP can be used repeatedly to further concentrate the discarded items in Ref. \cite{Pengpra}, and obtain a higher success probability.

In the concentration process, the atom state detection and photon state detection play prominent roles. Under current experimental conditions, both
the atom detection and single photon detection are imperfect with the detection efficiency $\eta<100\%$. Therefore, for comparing the success probability of our ECP with that in the ECP of Ref. \cite{Pengpra} under practical experimental conditions, it is necessary to consider the impact of the detection efficiency on the success probability. We suppose that the single photon detection efficiency and the single atom detection efficiency are $\eta_{p}$ and $\eta_{a}$, respectively. In the ECP of Ref. \cite{Pengpra}, as the single atom state and a pair of N-photon state need to be detected, its success probability can be written as
\begin{eqnarray}
P'_{total}&=&\eta_{p}^N\eta_{a}2|\alpha\beta|^{2},\label{p}
\end{eqnarray}
which shows an exponential decay with the photon number N. In our ECP, Alice only needs to detect the single atom state and the auxiliary single photon state in each concentration round. According to the description of Sec. III and Sec. IV, we can calculate the success probability in each concentration round as
\begin{eqnarray}
P_{1}&=&\eta_{a}\eta_{p}2|\alpha\beta|^{2},\nonumber\\
P_{2}&=&\eta_{a}\eta_{p}\frac{2|\alpha\beta|^{4}}{|\alpha|^{4}+|\beta|^{4}},\nonumber\\
P_{3}&=&\eta_{a}\eta_{p}\frac{2|\alpha\beta|^{8}}{(|\alpha|^{4}+|\beta|^{4})(|\alpha|^{8}+|\beta|^{8})},\nonumber\\
&\cdots\cdots&\nonumber\\
P_{K}&=&\eta_{a}\eta_{p}\frac{2|\alpha\beta|^{2^{K}}}{(|\alpha|^{4}+|\beta|^{4})(|\alpha|^{8}+|\beta|^{8})\cdots(|\alpha|^{2^{K}}+|\beta|^{2^{K}})^{2}}.\label{probability}
\end{eqnarray}
Therefore, the total success probability (P$_{total}$) of our ECP equals the sum of the success probability in each concentration round, which can be written as
\begin{eqnarray}
P_{total}=P_{1}+P_{2}+\cdots P_{K}=\eta_{a}\eta_{p}\sum\limits_{K=1}^{\infty} \frac{2|\alpha\beta|^{2^{K}}}{(|\alpha|^{4}+|\beta|^{4})(|\alpha|^{8}+|\beta|^{8})\cdots(|\alpha|^{2^{K}}+|\beta|^{2^{K}})^{2}},\label{pt}
\end{eqnarray}
which is independent of the photon number N.
\begin{figure}[!h]
\begin{center}
\includegraphics[width=8cm,angle=0]{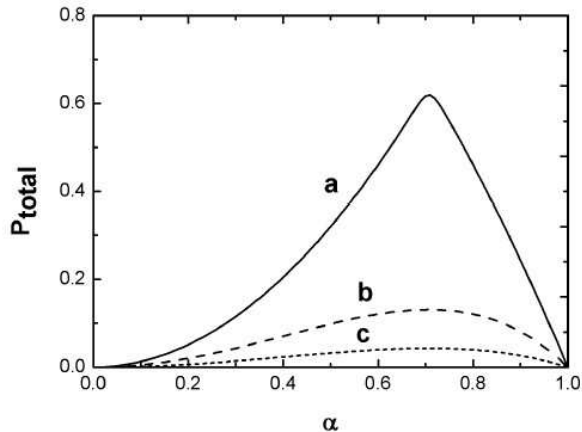}
\caption{The success probability ($P_{total}$) for both two ECPs under the imperfect detection, where the
efficiencies of both the photon state detection and atom state detection are set as $80\%$ for approximation. (a) The $P_{total}$ of our ECP being operated for 5 times. (b) The $P_{total}$ of the ECP in Ref. \cite{Pengpra}, when the photon number N is 5. (c) The $P_{total}$ of the ECP in Ref. \cite{Pengpra}, when the photon number N is 10. It can be seen under imperfect detection, the $P_{total}$ of the ECP in Ref. \cite{Pengpra} reduces largely with the photob number N, while the $P_{total}$ of our ECP can remains a relatively high level.}
\end{center}
\end{figure}

Recently, Heine \emph{et al.} reported their research result on the single atom detection. They have achieved $\eta_{a}= 66\%$ in the experiment \cite{atomdetection}. Moreover, they have shown that with some improvement, the single atom detection
 efficiency can achieve $\eta_{a}> 95\%$ in theory. The single photon detection has been a big difficulty under current experimental conditions, for
 the quantum decoherence effect of the photon detector \cite{photondetection}. In the optical range, $\eta_{p}$ is usually less than $30\%$ \cite{photondetection,photondetection1}. Lita \emph{et al.} reported their experimental result
 about the near-infrared single-photon detection. They showed the $\eta_{p}$ at 1556 nm can reach $95\%\pm 2\%$. Based on their research results, we make the numerical simulation on the total success probability (P$_{total}$) of both two ECPs. Fig. 4 shows the $P_{total}$ as a function of the entanglement coefficient $\alpha$. In Fig. 4, we assume $\eta_{p}=80\%$ and $\eta_{a}=80\%$ for approximation. In our ECP, we choose the repeating number $K=5$ (Fig. 4(a)), while in the ECP in Ref. \cite{Pengpra}, we choose the photon number $N=5$ (Fig. 4(b)) and $N=10$ (Fig. 4(c)). It is obvious that $P_{total}$ of the ECP in Ref. \cite{Pengpra} reduces largely with the increasing of the photon number N. Especially, according to Eq. (\ref{p}) and Eq. (\ref{pt}), when the photon number N is large, the $P'_{total}\rightarrow 0$, while our ECP can still get a relatively high success probability. Therefore, under practical experiment conditions, especially when the photon number N is large, our ECP shows greater advantage.

In summary, in the paper, we put forward an efficient ECP for arbitrary less-entangled multi-photon GHZ state with the help of the photonic Faraday rotation. Comparing with the ECP in Ref. \cite{Pengpra}, our ECP has some obvious advantages. First, our ECP only requires a pair of less-entangled multi-photon state and an auxiliary single photon, and can obtain the same success probability as the ECP in Ref. \cite{Pengpra}, which indicates our ECP is more economic. Second, our ECP only requires local operation, which can simplify the operation largely. Third, our ECP can be used repeatedly to further concentrate the discarded items in the ECP of Ref. \cite{Pengpra}, so that our ECP can obtain a higher success probability. Moreover, under practical imperfect detection conditions, the success probability of the ECP in Ref. \cite{Pengpra} shows an exponential decay with the photon number N, while our ECP can still obtain a relatively high success probability independent of N. Therefore, these features will make our ECP have a practical application in current quantum computation and communication.

\section*{ACKNOWLEDGEMENTS}
This work is supported by the National Natural Science Foundation of
China under Grant No. 11104159, and 61201164, Open Research
Fund Program of the State Key Laboratory of Low-Dimensional Quantum Physics Scientific, Tsinghua University,
Open Research Fund Program of National Laboratory of Solid State Microstructures under Grant No. M25020 and M25022,
 A Project Funded by the Priority Academic Program Development of Jiangsu Higher Education Institutions.


\begin{thebibliography}{99}

\bibitem{rmp} N. Gisin, G. Ribordy, W. Tittel, and H. Zbinden, Rev. Mod. Phys. \textbf{74}, 145 (2002).

\bibitem{teleportation} C. H. Bennett, G. Brassard, C. Crepeau, R. Jozsa, A. Peres, and W. K. Wootters, Phys. Rev. Lett. \textbf{70}, 1895 (1993).

\bibitem{cteleportation} A. Karlsson and M. Bourennane, Phys. Rev. A \textbf{58}, 4394 (1998); F. G. Deng, C. Y. Li, Y. S.
Li, H. Y. Zhou, and Y. Wang, Phys. Rev. A \textbf{72}, 022338 (2005).

\bibitem{densecoding} C. H. Bennett and S. J. Wiesner, Phys. Rev. Lett. \textbf{69}, 2881 (1992).

\bibitem{densecoding1} X. S. Liu, G. L. Long, D. M. Tong, and L. Feng, Phys. Rev. A \textbf{65}, 022304 (2002).

\bibitem{densecoding2} C. Wang, F. G. Deng, Y. S. Li, X. S. Liu, and G. L. Long, Phys. Rev. A \textbf{71}, 044305 (2005).

\bibitem{QSDC} F. G. Deng, G. L. Long, and X. S. Liu, Phys. Rev. A \textbf{68}, 042317 (2003).
\bibitem{QSDC1} G. L. Long and X. S. Liu, Phys. Rev. A \textbf{65}, 032302 (2002).

\bibitem{QSDC2} C. Wang, F. G. Deng, Y. S. Li, X. S. Liu, and G. L. Long, Phys. Rev. A \textbf{71}, 044305 (2005).

\bibitem{Ekert91} A. K. Ekert, Phys. Rev. Lett. \textbf{67}, 661 (1991).

\bibitem{QKDdeng1} F. G. Deng and G. L. Long, Phys. Rev. A \textbf{68}, 042315 (2003).

\bibitem{QKDdeng2} X. H. Li, F. G. Deng, and H. Y. Zhou, Phys. Rev. A \textbf{78}, 022321 (2008).

\bibitem{QSTS} F. G. Deng, X. H. Li, C. Y. Li, P. Zhou, and H. Y. Zhou, Phys. Rev. A \textbf{72}, 044301 (2005).

\bibitem{QSS1} M. Hillery, V. Bu$\breve{z}$ek, and A. Berthiaume, Phys. Rev. A \textbf{59}, 1829 (1999).


\bibitem{QSS2} A. Karlsson, M. Koashi, and N. Imoto, Phys. Rev. A \textbf{59}, 162 (1999).
\bibitem{QSS3} L. Xiao, G. L. Long, F. G. Deng, and J. W. Pan, Phys. Rev. A \textbf{69}, 052307 (2004).

\bibitem{memory} L. M. Duan, M. D. Lukin, J. I. Cirac, and P. Zoller, Nature \textbf{414}, 413 (2001).

\bibitem{C.H.Bennett1} C. H. Bennett, G. Brassard, S. Popescu, B. Schumacher, J. A. Smolin, and W. K. Wootters, Phys. Rev. Lett. \textbf{76}, 722 (1996).

\bibitem{Pan1} J. W. Pan, C. Simon, and A. Zellinger, Nature (London) \textbf{410}, 1067 (2001).

\bibitem{C.H.Bennett2} C. H. Bennett, H. J. Bernstein, S. Popescu, and
B. Schumacher, Phys. Rev. A \textbf{53}, 2046 (1996).

\bibitem{swapping1} S. Bose, V. Vedral, and P. L. Knight, Phys. Rev A \textbf{60}, 194 (1999).
\bibitem{swapping2} B. S. Shi, Y. K. Jiang, and G. C. Guo, Phys. Rev. A \textbf{62}, 054301 (2000).

\bibitem{Yamamoto1} T. Yamamoto, M. Koashi, and N. Imoto, Phys. Rev. A \textbf{64}, 012304 (2001).
\bibitem{zhao1} Z. Zhao, J. W. Pan, and M. S. Zhan, Phys. Rev. A \textbf{64}, 014301 (2001).


\bibitem{shengpra2} Y. B. Sheng, F. G. Deng, and H. Y. Zhou, Phys. Rev. A \textbf{77}, 062325 (2008).
\bibitem{shengqic} Y. B. Sheng, F. G. Deng, and H. Y. Zhou, Quant. Inf. Comput. \textbf{10}, 272 (2010).
\bibitem{shengpla} Y. B. Sheng, F. G. Deng, and H. Y. Zhou, Phys. Lett. A \textbf{373}, 1823 (2009).

\bibitem{dengpra} F. G. Deng, Phys. Rev. A \textbf{85}, 022311 (2012).
\bibitem{shengpra3} Y. B. Sheng, L. Zhou, S. M. Zhao, and B. Y. Zheng, Phys. Rev. A \textbf{85}, 012307 (2012).

\bibitem{shengpra4} Y. B. Sheng, L. Zhou, and S. M. Zhao, Phys. Rev. A \textbf{85}, 044305 (2012).

\bibitem{wangxb} X. B. Wang and H. Fan, Phys. Rev. A \textbf{68}, 060302 (2003).

\bibitem{wangc} C. Wang, Y. Zhang, and G. S. Jin, Phys. Rev. A \textbf{84}, 032307 (2011).

\bibitem{Pengpra} Z. H. Peng, J. Zou, X. J. Liu, Y. J. Xiao, and L. M. Kuang, Phys. Rev. A \textbf{86}, 034305 (2012).
\bibitem{cavity} J. I. Cirac, P. Zoller, H. J. Kimble, and H. Mabuchi, Phys. Rev. Lett. \textbf{78}, 3221 (1997).

\bibitem{cavity1} J. I. Cirac, A. K. Ekert, S. F. Huelga, and C. Macchiavello, Phys. Rev. A \textbf{59}, 4249 (1999).
\bibitem{cavity2} Q. A. Turchette, C. J. Hood, W. Lange, H. Mabuchi, and H. J. Kimble, Phys. Rev. Lett. \textbf{75}, 4710 (1995).
\bibitem{cavity3}  M. Brune, E. Hagley, J. Dreyer, X. Maitre, A. Maali, C. Wunderlich, J. M. Raimond, and S. Haroche, Phys. Rev.
Lett. \textbf{77}, 4887 (1996).

\bibitem{cavity4} K. Mattle, H. Weinfurter, P. G. Kwiat, and A. Zeilinger, Phys. Rev. Lett. \textbf{76}, 4656 (1996).
\bibitem{cavity5} P. Xue and Y. F. Xiao, Phys. Rev. Lett. \textbf{97}, 140501 (2006).
\bibitem{cavity6} L. M. Duan and H. J. Kimble, Phys. Rev. Lett. \textbf{92}, 127902 (2004).
\bibitem{cavity7} Y. F. Xiao, X. M. Lin, J. Gao, Y. Yang, Z. F. Han, and G. C. Guo, Phys. Rev. A \textbf{70}, 042314 (2004).
\bibitem{cavity8} J. Cho and H. W. Lee, Phys. Rev. Lett. \textbf{95}, 160501 (2005).
\bibitem{cavity9} L. M. Duan, B. Wang, and H. J. Kimble, Phys. Rev. A \textbf{72}, 032333 (2005).

\bibitem{node} X. M. Lin, Z. W. Zhou, M. Y. Ye, Y. F. Xiao, and G. C. Guo, Phys. Rev. A \textbf{73}, 012323 (2006).
\bibitem{node1} Z. J. Deng, X. L. Zhang, H. Wei, K. L. Gao, and M. Feng, Phys. Rev. A \textbf{76}, 044305 (2007).
\bibitem{node2} H. Wei, Z. J. Deng, X. L. Zhang, and M. Feng, Phys. Rev. A \textbf{76}, 054304 (2007).

\bibitem{MTR} B. Dayan, A. S. Parkins, T. Aoki, E. P. Ostby, K. I. Vahala, and
H. J. Kimble, Science \textbf{319}, 1062 (2008).

\bibitem{Faraday} B. Julsgaard, A. Kozhekin, and E. S. Polzik, Nature (London). \textbf{413}, 400 (2001).

\bibitem{faraday3} D. F. Walls and G. J. Milburn, Quantum Optics. Springer, Berlin (1994).

\bibitem{faraday1} H. J. Carmichael, Springer, Berlin (2008).
\bibitem{r} J. H. An, M. Feng, C. H. Oh, Phys. Rev. A \textbf{79}, 032303 (2009).
\bibitem{atomdetection} D. Heine, W. Rohringer, D. Fischer, M. Wilzbach, T. Raub, S. Loziczky, L. X. Yuan,
S. Groth, B. Hessmo, and J. Schmiedmayer, New J. Phys. \textbf{12}, 095005 (2010).

\bibitem{photondetection} V. D'Auria, N. Lee, T. Amri, C. Fabre, J. Laurat, Phys. Rev. Lett. \textbf{107}, 050504 (2011).
\bibitem{photondetection1} D. Henrich, L. Rehm, S. $D\ddot{o}rner$, M. Hofherr, K. Ilin, A. Semenov, M. Siegel, arXiv:1210.3988 (2012).
\bibitem{photondetection2} A. E. Lita, A. J. Miller, and S. W. Nam, Optics Express, \textbf{16}, 3032 (2008).

\end{thebibliography}
\end{document}